\documentclass{aa} 
\usepackage[varg]{txfonts}
\usepackage[utf8]{inputenc}
\usepackage{natbib}
\bibpunct{(}{)}{;}{a}{}{,}% to follow the A&A style
\usepackage{graphicx}
\usepackage{amsmath}
\usepackage{booktabs}
\usepackage{float}
\usepackage{makecell}
\usepackage{siunitx}
\usepackage{afterpage}
\usepackage{color}
\usepackage{hyperref}
\hypersetup{
        colorlinks=true,% color of internal links
        breaklinks=true,%
        linkcolor=blue,% color of internal links
        citecolor=blue,% color of links to bibliography
        filecolor=blue,% color of file links
        urlcolor=blue,%
        unicode=false,% non-Latin characters in Acrobat bookmarks
        pdftoolbar=true,% show Acrobat toolbar?
        pdfmenubar=true,% show Acrobat menu?
        pdffitwindow=false,% window fit to page when opened
        pdfstartview={Fit},% fits the width of the page to the window
        pdftitle={3D shape of OrionA},%title
        pdfauthor={Josefa},% author
        pdfsubject={OrionA},%
        pdfcreator={Josefa},% creator of the document
        pdfnewwindow=true,% links in new window
        pdfdisplaydoctitle=true%display document title instead of file name in title b
}

\begin{document}

\title{3D shape of Orion\,A from \textit{Gaia} DR2}

\titlerunning{3D shape of Orion\,A from \textit{Gaia} DR2}

\author{Josefa~E.~Gro\ss schedl\inst{1}
          \and
          Jo\~ao~Alves\inst{1}
          \and
          Stefan~Meingast\inst{1}
          \and 
          Christine~Ackerl\inst{1}
          \and
          Joana~Ascenso\inst{2}
          \and
          Herv\'e~Bouy\inst{3} 
          \and
          Andreas~Burkert\inst{4,5}
          \and
          Jan~Forbrich\inst{6,7} 
          \and
          Verena~F\"urnkranz\inst{1}
          \and
          Alyssa~Goodman\inst{7}
          \and
          \'Alvaro~Hacar\inst{8} 
          \and
          Gabor~Herbst-Kiss\inst{1}
          \and 
          Charles~J.~Lada\inst{7}
          \and 
          Irati~Larreina\inst{1}
          \and
          Kieran~Leschinski\inst{1}
          \and
          Marco~Lombardi\inst{9} 
          \and
          Andr\'e Moitinho\inst{10}
          \and
          Daniel~Mortimer\inst{11}
          \and
          Eleonora~Zari\inst{8}
        }%
  
   \authorrunning{J.~Gro\ss schedl}

   \institute{Universit\"at Wien, Institut f\"ur Astrophysik, T\"urkenschanzstrasse 17, 1180 Wien, Austria, \\
            \email{josefa.elisabeth.grossschedl@univie.ac.at}
        \and
            Universidade do Porto, Departamento de Engenharia F\'isica da Faculdade de Engenharia, Rua Dr.\,Roberto Frias, P-4200-465, Porto, Portugal
        \and
            Laboratoire d'Astrophysique de Bordeaux, Universite de Bordeaux, All\'ee Geoffroy Saint-Hilaire, CS 50023, 33615 PESSAC CEDEX, France
        \and
            Universit\"ats-Sternwarte Ludwig-Maximilians-Universit\"at (USM), Scheinerstr. 1, 81679 M\"unchen, Germany
        \and
            Max-Planck-Fellow, Max-Planck-Institut f\"ur extraterrestrische Physik (MPE), Giessenbachstr. 1, 85748 Garching, Germany
        \and
            Centre for Astrophysics Research, School of Physics, Astronomy and Mathematics, University of Hertfordshire, College Lane, Hatfield AL10 9AB, UK
        \and
            Harvard-Smithsonian Center for Astrophysics, 60 Garden Street, Cambridge, MA 02138, USA
        \and
            Leiden Observatory, Niels Bohrweg 2, 2333 CA Leiden, The Netherlands  
        \and
            University of Milan, Department of Physics, via Celoria 16, 20133 Milan, Italy
        \and
            CENTRA, Universidade de Lisboa, FCUL, Campo Grande, Edif. C8, 1749-016 Lisboa, Portugal
        \and
            Cavendish Laboratory, University of Cambridge, 19 J.J. Thomson Avenue, Cambridge, CB3 0HE, UK
             }%

% \date{Received July 19}

\abstract{We use the \textit{Gaia} DR2 distances of about 700 mid-infrared selected young stellar objects in the benchmark giant molecular cloud Orion\,A to infer its 3D shape and orientation. We find that Orion\,A is not the fairly straight filamentary cloud that we see in (2D) projection, but instead a cometary-like cloud oriented toward the Galactic plane, with two distinct components: a denser and enhanced star-forming (bent) Head, and a lower density and star-formation quieter $\sim$75 pc long Tail. 
The true extent of Orion\,A is not the projected $\sim$40\,pc but $\sim$90\,pc, making it by far the largest molecular cloud in the local neighborhood. Its aspect ratio ($\sim$30:1) and high column-density fraction ($\sim$45\%) make it similar to large-scale Milky Way filaments (``bones''), despite its distance to the galactic mid-plane being an order of magnitude larger than typically found for these structures.}


\maketitle

\section{Introduction} \label{Intro}

\defcitealias{Schlafly2014}{Schlafly et al.}

The archetypal giant molecular cloud (GMC) Orion\,A is the most active star-forming region in the local neighborhood, having spawned $\sim$3,000 young stellar objects (YSOs) in the last few million years \citep[e.g.,][]{Megeath2012, Furlan2016, Grossschedl2018}. Some of the most basic observables of the star-formation process, including star-formation rates and history, age spreads, multiplicity, the initial mass function, and protoplanetary disk populations, have been derived for this benchmark region \citep[e.g.,][]{Bally2008,Muench2008}. Previous distance estimates to the Orion Nebula Cluster (ONC), the richest cluster toward the northern end of the cloud, put this object at about \SI{400}{pc} from Earth \citep[e.g.,][]{Sandstrom2007, Menten2007, Hirota2007, Kim2008, Kuhn2018}. Moreover, there has been some evidence that the northern part of the cloud, including the ONC (or ``Head''), is closer than the southern part (or ``Tail'')\footnote{For simplicity we classify the Orion\,A cloud roughly into Head and Tail; the Tail represents the less star-forming part.}, containing L1641 and L1647 \citep{Schlafly2014, Kounkel2017, Kounkel2018}.

\begin{table*}[!ht] 
\begin{center}
\small
\caption{Distances to sub-regions in Orion\,A from the Literature.} 
\renewcommand{\arraystretch}{1.2}
\begin{tabular}{llll}
\hline \hline
\multicolumn{1}{c}{Reference} & \multicolumn{1}{c}{Method} & \multicolumn{1}{c}{Region} & \multicolumn{1}{c}{Distance (pc)}  \\
 & & & \multicolumn{1}{c}{(pc)}  \\
\hline

\citet{Genzel1981} & proper motion and radial velocity & Orion KL & $480\pm80$ \\
 & of H$_2$O masers &  &  \\
\citet{Hirota2007} & VERA/VLBI & Orion KL & $437\pm19$ \\
\citet{Menten2007} & VLBI & ONC & $414\pm7$ \\
\citet{Sandstrom2007} & VLBI & ONC & $389^{+24}_{-21}$ \\
\citet{Kim2008} & VERA/VLBI & Orion KL & $418\pm6$ \\
\citet{Lombardi2011} & density of foreground stars & Orion\,A & $371\pm10$ \\
\hline
\citet{Schlafly2014}\tablefootmark{a} & PanSTARRS optical reddening  & 
      $(l/b)$ at $(\SI{208.4}{\degree},\SI{-19.6}{\degree})$ north of the ONC & $418^{+43}_{-34}$ \\
 & \citep{Green2014} & $(l/b)$ at $(\SI{209.1}{\degree},\SI{-19.9}{\degree})$ west of the ONC & $478^{+84}_{-59}$ \\
 &  & $(l/b)$ at $(\SI{209.0}{\degree},\SI{-20.1}{\degree})$ west of the ONC & $416^{+42}_{-36}$ \\
 &  & $(l/b)$ at $(\SI{209.8}{\degree},\SI{-19.5}{\degree})$ north to L1641-North & $580^{+161}_{-107}$ \\
 &  & $(l/b)$ at $(\SI{212.2}{\degree},\SI{-18.6}{\degree})$ east to L1641-South & $490^{+27}_{-27}$ \\
 &  & $(l/b)$ at $(\SI{212.4}{\degree},\SI{-19.9}{\degree})$ west to L1641-South & $517^{+44}_{-38}$ \\
 &  & $(l/b)$ at $(\SI{214.7}{\degree},\SI{-19.0}{\degree})$ south-east of L1647-South & $497^{+42}_{-36}$ \\
 \hline
\citet{Kounkel2017}\tablefootmark{a} & VLBI & 15 YSOs near the ONC & $388\pm5$ \\
 &  & 2 YSOs near L1641-South & $428\pm10$ \\
 \hline
\citet{Kounkel2018} & {\it Gaia} DR2 of APOGEE-2 sources & ONC & $386\pm3$ \\
 & + HR-diagram selection  & L1641-South & $417\pm4$\ \\
 &  & L1647 & $443\pm5$ \\
 \hline
\citet{Kuhn2018} & {\it Gaia} DR2 of {\it Chandra} X-ray sources & ONC & $403^{+7}_{-6}$ \\
 &  & north and south to ONC & $\sim395$ \\
 \hline
\end{tabular}
\renewcommand{\arraystretch}{1}
\label{tab:literature}
\tablefoot{
\tablefoottext{a}{See also Fig.~\ref{fig:scatter_literature}.}
}
\end{center}
\end{table*}

To know the true 3D spatial shape and orientation of this giant filamentary structure would allow one not only to determine accurate cloud and YSO masses, luminosities, and separations for this benchmark region, but it would also bring important hints on the formation of GMCs in the disk of the Milky Way. \citet{Schlafly2014} first found an indication of a distance gradient across Orion\,A (see Table~\ref{tab:literature}), using a method based on optical reddening of stars \citep{Green2014} which is not sensitive to regions of high column-density. \citetalias{Schlafly2014} found that the Tail of the cloud is about 70\,pc more distant than the ONC region. \citet{Kounkel2017} conducted 15 VLBI observations toward young stars near the ONC, and two observations toward L1641-South. These observations again suggest an inclination of the cloud away from the plane of the sky, with a difference in distance of about 40\,pc from Head to Tail (until L1641-South). The distances reported in \citet{Schlafly2014} and \citet{Kounkel2017} are presented in Fig.~\ref{fig:scatter_literature} and in Table~\ref{tab:literature}. \citet{Kounkel2018} continued the analysis of this region by using new APOGEE-2 data combined with the newly released {\it Gaia} DR2 catalog \citep{Brown2018}. In this recent paper, they focus on stellar populations and the star-formation history across the whole Orion complex in a high dimensional space using a clustering algorithm. They report a more distant Tail compared to the Head (about 55\,pc distance difference).

In this paper we have used the newly released {\it Gaia} DR2 to infer the 3D shape and orientation of Orion\,A. As a proxy to the cloud distance we will use the latest catalog of mid-infrared selected YSOs in this cloud, with ages $\lesssim 3$ Myr, for which a {\it Gaia} DR2 parallax measurement exists. These very young stars lie relatively close to, or are still embedded in the Orion\,A GMC, sharing the same velocity as the cloud \citep{Tobin2009, Hacar2016}, and are thus the best tracer of the cloud's shape.

\begin{figure}[!ht]
    \centering
    \includegraphics[width=1\linewidth]{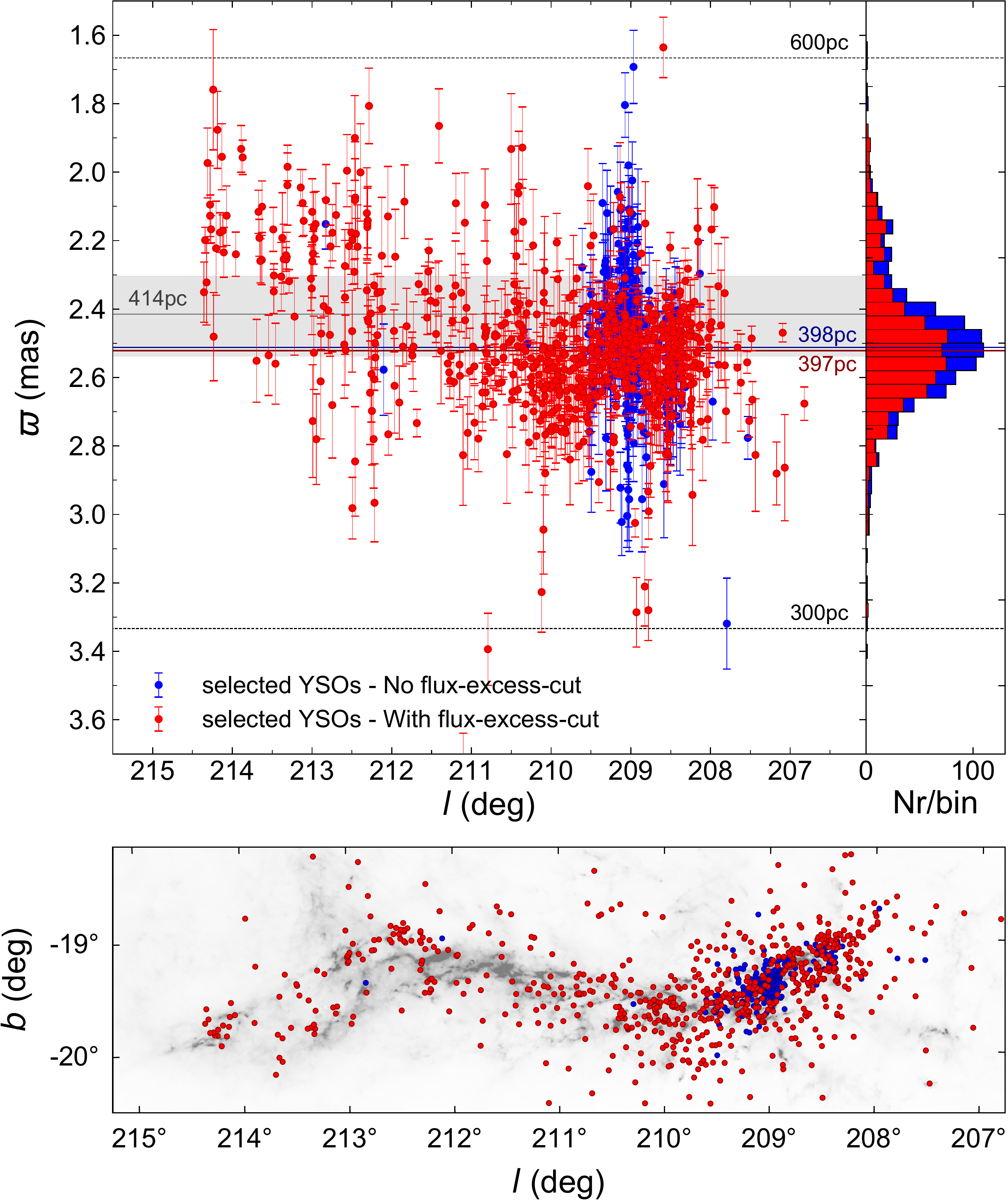}
    \caption{
    {\it Gaia} DR2 $\varpi$ of YSOs with IR-excess in Orion\,A versus $l$ (top, $\sigma_\varpi$ as error-bars), and projected YSO distribution displayed on the {\it Herschel} map (bottom). Red are YSOs that pass the applied selection criteria as discussed in the first two steps in Sect.~\ref{Data}. The blue sources represent the sources lost when the flux-excess-cut is applied. This highlights that nebulae (near the ONC, see map) cause additional $\varpi$ uncertainties, not reflected in $\sigma_\varpi$.
    The 1D distribution of $\varpi$ for both samples is shown in the histogram on the right. The red and blue middle lines show the median $\varpi$ of the samples. 
    The lower and upper borders (black dashed lines) indicate the applied distance cuts to avoid possible foreground or background contamination when deducing the average distances.
    The middle gray line shows the distance to the ONC of 414\,pc \citep{Menten2007}, while the gray shaded band represents the 2D projected size of the cloud of about 40\,pc at 414\,pc.
    }
    \label{fig:scatter}
\end{figure}

\begin{figure}[!ht]
    \centering
    \includegraphics[width=0.95\linewidth]{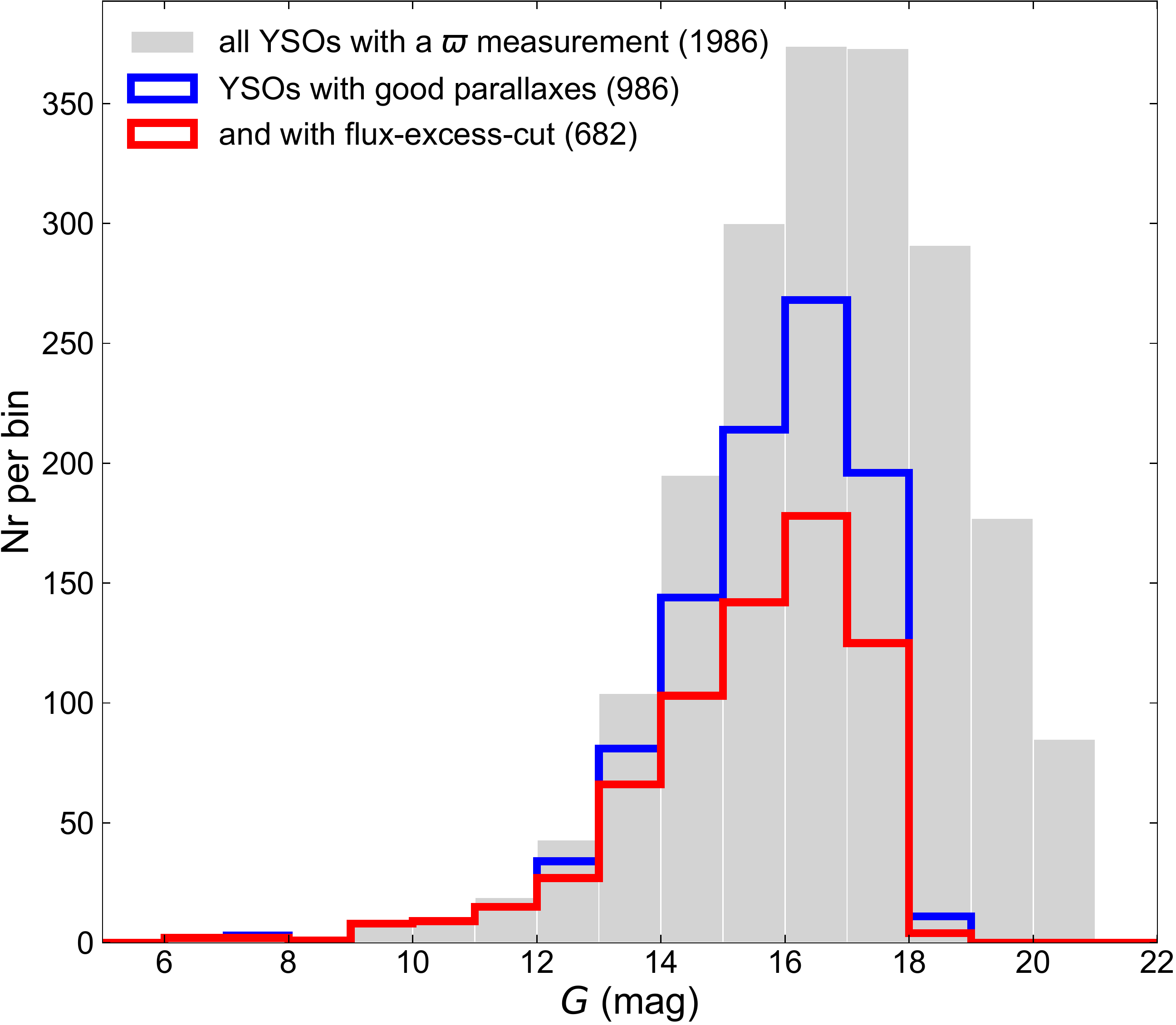}
    \caption{Histogram of {\it Gaia} DR2 $G$-band magnitudes. The gray distribution shows all YSOs toward Orion\,A with measured {\it Gaia} DR2 parallaxes. The red and blue distributions show the YSO samples that pass our required selection criteria, while we distinguish sources with (red) and without (blue) flux-excess-cut (see also Fig.~\ref{fig:scatter}).}
    \label{fig:mag_hist}
\end{figure}

\begin{figure*}[!ht]
    \centering
    \includegraphics[width=0.95\linewidth]{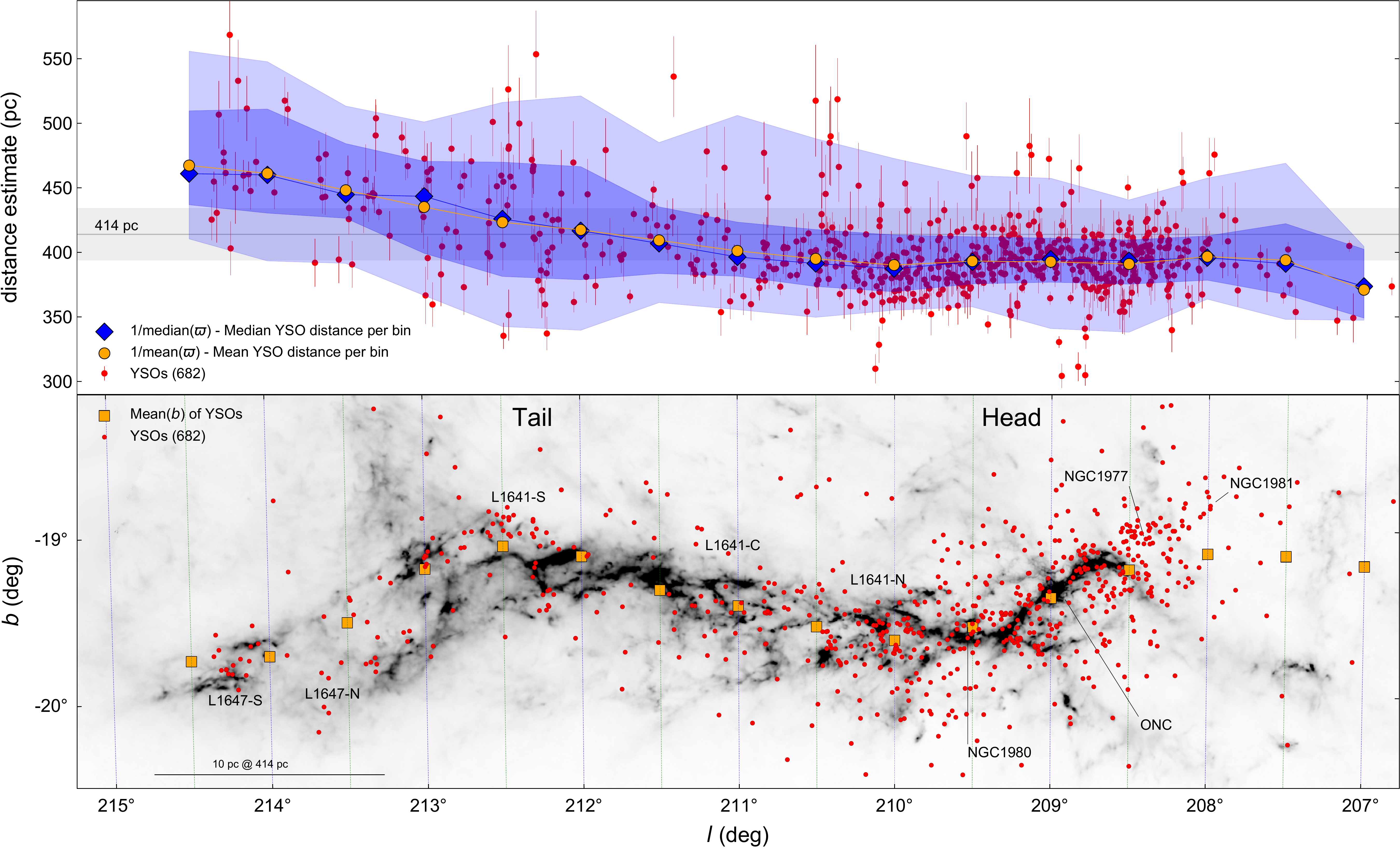}
    \caption{
    Top: Distance estimates ($1/\varpi$) for YSOs in Orion\,A versus $l$ and their average distances per $\Delta l$. YSOs are shown as red dots with error-bars ($\sigma_\varpi/\varpi^2$). Over-plotted are the median (blue diamonds) and mean (orange circles) distances per $\Delta l = \SI{1}{\degree}$ (blue/green vertical lines on the bottom map correspond to the bin boundaries, factor two over-sampled). The 1$\sigma$ and 2$\sigma$ lower and upper percentiles are shown as blue shaded areas. The horizontal gray line represents the \citet{Menten2007} distance to the ONC at $\SI{414}{pc}$ with a range of $\SI{\pm 20}{pc}$ (gray shaded area) corresponding to the projected extent in $l$ of the cloud ($\SI{\sim40}{pc}$ at $\SI{414}{pc}$).
    Bottom: Distribution of the YSOs in Galactic coordinates projected on the {\it Herschel map}. The displayed area corresponds to the VISTA observed region \citep{Meingast2016}. The dark shades of the gray scale indicate regions of high dust column-density (or high extinction). The distribution of YSOs follows the high density regions of the cloud, shown by their mean($b$) positions per $\Delta l$ (orange squares).
   }
    \label{fig:mean}
\end{figure*}

\section{Observations and data selection} \label{Data}

\begin{figure}[!ht]
    \centering
    \includegraphics[width=0.9\linewidth]{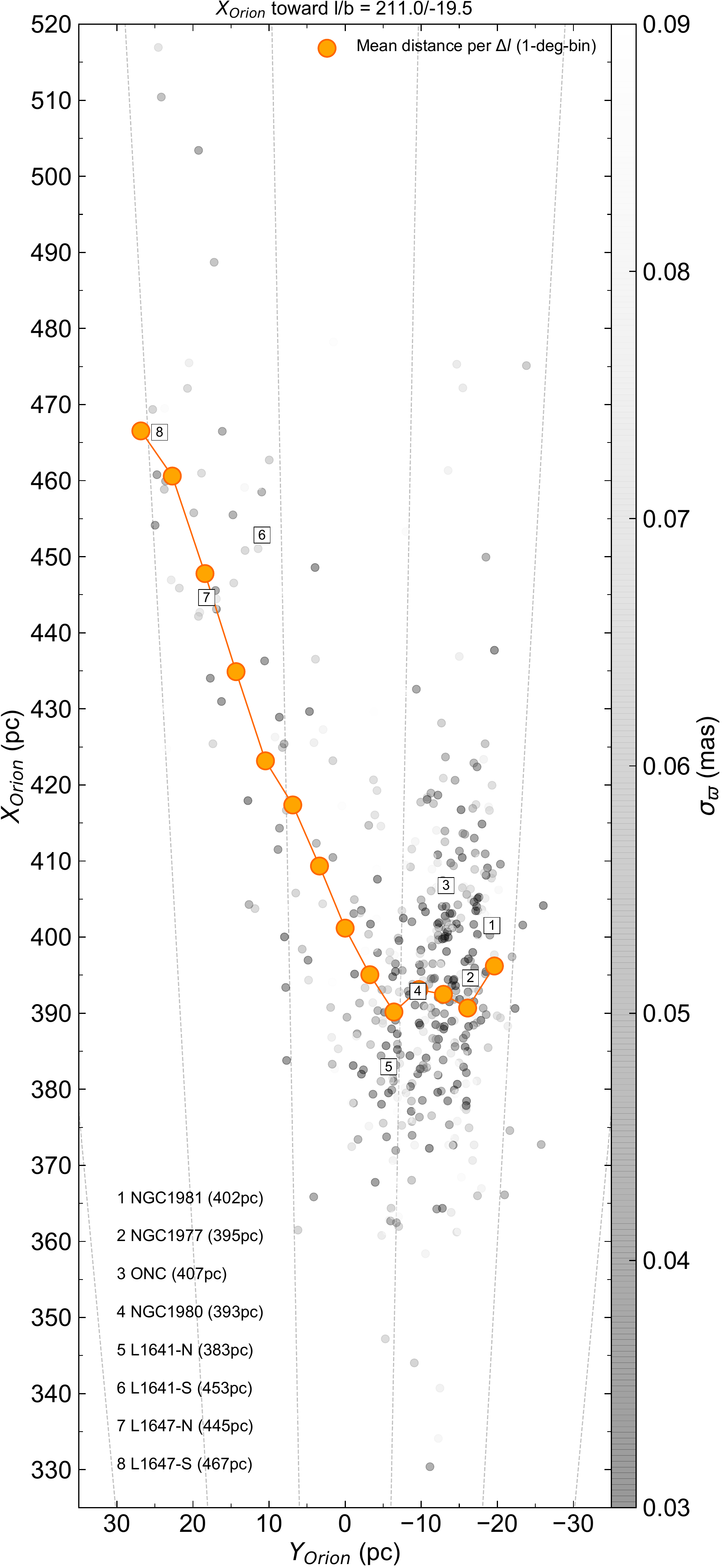}
    \caption{
    YSO distribution and YSO's mean positions projected in a cartesian plane. In this coordinate frame the Sun (at $X,Y,Z=0,0,0$) is connected to the location of Orion\,A with $X_{Orion}$ pointing toward $(l/b) = (\SI{211}{\degree}/\SI{-19.5}{\degree})$. Consequently, $X_{Orion}$ is similar to the distance from the Sun, while the $Y_{Orion}$  and $Z_{Orion}$ components coincide roughly with the $l$ and $b$ distribution, respectively.
    The YSOs are shown in gray scale colored by $\sigma_\varpi$.
    The mean positions per $\Delta l$ are shown as orange filled circles (as in Fig.~\ref{fig:mean}), while the two rightmost points from Fig.~\ref{fig:mean} are excluded. The gray dashed lines are lines of constant $l$ as viewed from the Sun ($\SI{2}{\degree}$ steps from $l = \SI{206}{\degree}$ to $\SI{216}{\degree}$). For orientation, the numbered boxes show the mean positions of YSOs projected near eight known clusters, as listed in the bottom left legend. In brackets we give the estimated distances (derived from {\it Gaia} DR2 parallaxes), which are used to plot the boxes.
    }
    \label{fig:xy_stretch}
\end{figure}

\begin{figure*}[!ht]
    \centering
    \includegraphics[width=0.95\linewidth]{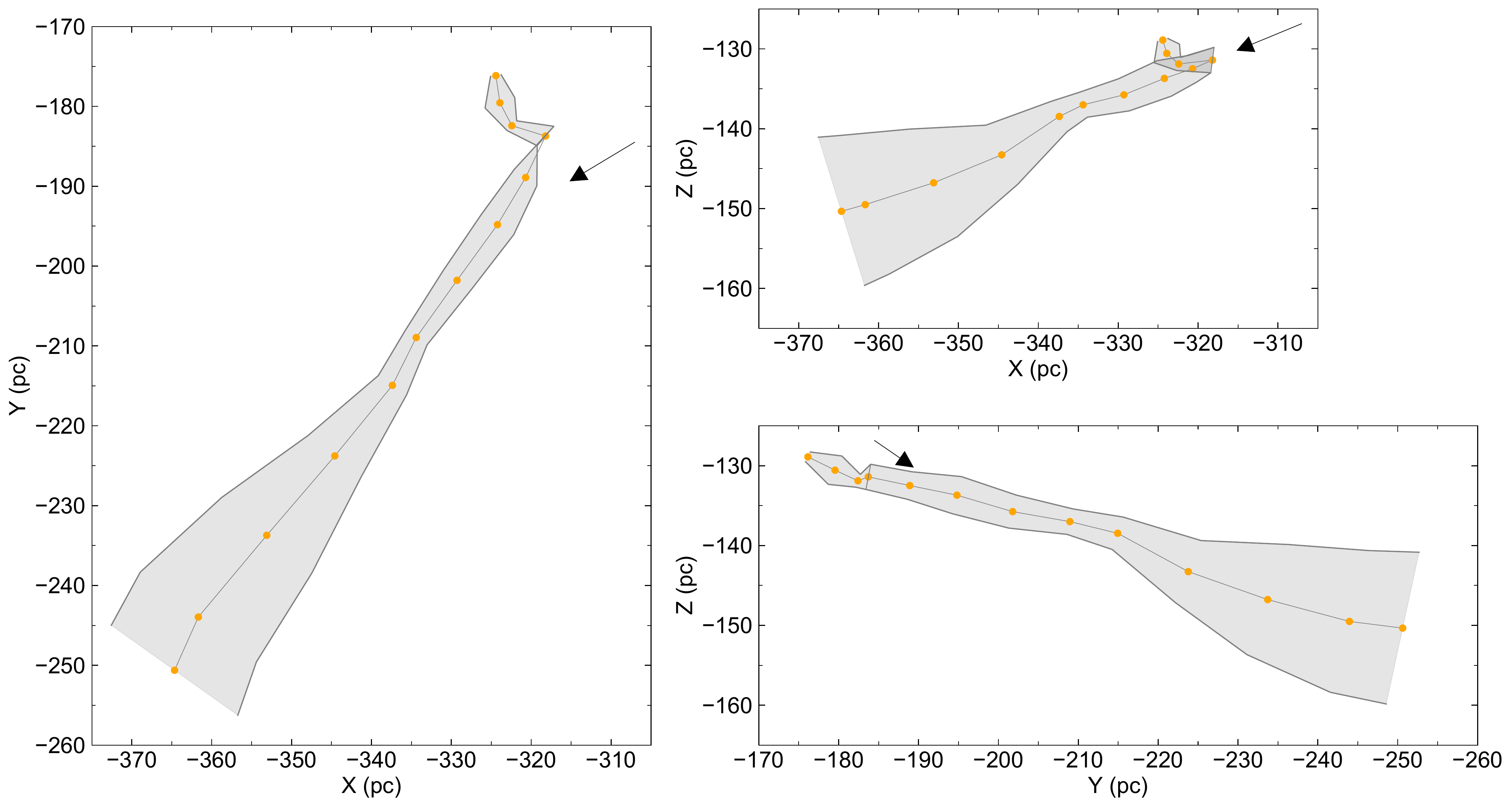}
    \caption{
    3D orientation of the Orion\,A GMC in Galactic cartesian coordinates ($X$ positive toward the Galactic center, $Y$ positive toward the Galactic east, $Z$ positive toward the Galactic north). The orange dots represent the mean positions of YSOs per $\Delta l$ (see Fig.~\ref{fig:mean}), while only using those on top of high column-density. The gray shaded area shows an idealized 3D cloud shape in each projection at $A_\mathrm{K,Herschel}\gtrsim\SI{0.57}{mag}$ ($A_V\gtrsim\SI{5}{mag}$), assuming a symmetric cylindrical shape, meaning that the filament is as deep as it is wide in the sky. For orientation, the black arrows indicate the line-of-sight from the Sun. Each arrow points toward $(l,b) = (211.0, -19.5)$ plotted from $d=$ 380\,pc to 390\,pc. 
    }
    \label{fig:xyz}
\end{figure*}

We have used the Orion\,A YSO catalog of \citet{Grossschedl2018} which revisited the catalog of \citet{Megeath2012} \citep[including updates from][]{Megeath2016, Furlan2016, Lewis2016}, and added about 200 new YSO candidates from a dedicated ESO-VISTA near-infrared survey covering the whole Orion\,A region \citep[$\SI{\sim18}{deg\square}$,][]{Meingast2016}, making it the most complete (2D) distribution of YSOs toward this cloud. The catalog contains 2,978 YSO candidates with IR-excess, classified into 2,607 pre-main-sequence stars with disks (Class\,II), 183 flat-spectrum sources, and 188 protostars (Class\,0/I). The on-sky distribution of these sources generally follows the high density regions of the cloud (see Fig.~\ref{fig:mean}, bottom). Combined with their youth, this makes them a good proxy for cloud distances.

To infer the distance along the Orion\,A GMC we averaged over YSO's parallaxes ($\varpi$) in equally sized bins of Galactic longitude ($\Delta l$). To derive distances from parallaxes we have investigated both the inverse of $\varpi$ and the Bayesian distance estimates from \citet{BailerJones2018}, which account for the non-linearity of the transformation parallax to distance. At a distance of 400\,pc, the mean difference between the two methods is about 1\% for DR2, meaning that the final result in this paper will be virtually independent of the method used to infer distances. 
Moreover, we do not include a global parallax offset of 0.029\,mas or 0.08\,mas, as discussed in \citet{Lindegren2018} and \citet{Stassun2018}, since it is very uncertain how or if this effect influences the parallaxes of our sample toward Orion\,A. Besides, the presence of a small offset does not affect the result in this paper. To summarize, in this paper we use parallaxes when possible, and only then we derive the mean or median distance from the inverse of the $\mathrm{mean}(\varpi)$ or $\mathrm{median}(\varpi)$.

Before cross-matching the YSO sample with \textit{Gaia} DR2 data we checked the effect of proper motions on the cross-match. To this end, we transformed \textit{Gaia} J2015.5 coordinates into J2000. The effect toward Orion\,A is marginal, with a mean separation between J2015.5 and J2000 of $\SI{0.09}{\arcsec}$, smaller than the astrometric accuracy of the VISTA survey (observed in 2013). We used then a $\SI{1}{\arcsec}$ cross-match radius between the original {\it Gaia} J2015.5 and VISTA coordinates. This results in 1,986 cross-matches of DR2 parallaxes with IR-excess YSOs (67\% of the original YSO catalog).

Since we are interested in reliable anchor points along the cloud, and given the relatively good statistics, we chose a conservative selection criteria for the final sample \citep[informed by][]{Babusiaux2018, Lindegren2018, Arenou2018, Evans2018}, which is described in the following three steps. In a first step we applied several cuts to get reliable parallax measurements:\footnote{Shortcuts for {\it Gaia} parameters: \\
$\varpi$: {\tt parallax} [mas] \\
$\sigma_\varpi$: {\tt parallax\_error} [mas] \\
$G$: {\tt phot\_g\_mean\_mag} [mag]\\
$\sigma_G$: 1.0857 / {\tt phot\_g\_mean\_flux\_over\_error} [mag]\\
$\epsilon_i$: {\tt astrometric\_excess\_noise} [mas]\\
$\mathrm{sig}\_\epsilon_i$: {\tt astrometric\_excess\_noise\_sig} (significance)
}
\begin{equation}
\begin{aligned}
& \sigma_\varpi/\varpi < 0.1 \\ 
& \sigma_G < \SI{0.03}{mag} \\
& \mathtt{astrometric\_sigma5d\_max} < \SI{0.3}{mas} \\
& \mathtt{visibility\_periods\_used} > 8 \\
& (\epsilon_i \leq \SI{1}{mas}) \,\,\,\mathrm{OR}\,\,\, (\epsilon_i > \SI{1}{mas} \,\,\,\mathrm{AND}\,\,\, \mathrm{sig}\_\epsilon_i \leq 2) \\
\end{aligned}
\end{equation}
Bright nebulosities and crowded regions can cause further uncertainties, which especially effect the ONC region. Hence, in a second step, we excluded sources showing a flux excess\footnote{using the ratio of the fluxes $(I_\mathrm{BP}-I_\mathrm{RP})/I_\mathrm{G}$:\\
$\mathtt{phot\_bp\_rp\_excess\_factor}$}, by applying the following flux-excess-cut, similar to \citet{Evans2018}:
\begin{equation}
(I_\mathrm{BP}-I_\mathrm{RP})/I_\mathrm{G} > 1.35 + 0.06 (G_\mathrm{BP}-G_\mathrm{RP})^2
\end{equation}
This condition significantly reduces the distance scatter near the ONC (see Fig.~\ref{fig:scatter}), 
but it does not significantly affect the averaged parallaxes along the cloud, since the scatter is more or less symmetric.
Finally, in a third step, we used only sources in a distance interval of $\SI{300}{pc} < d < \SI{600}{pc}$ (or $\SI{3.333}{mas} \gtrsim \varpi \lesssim \SI{1.666}{mas}$), since an examination of the parallax distribution (Fig.~\ref{fig:scatter}) shows a clear drop in density of sources beyond these boundaries. This prevents the contamination by outliers when averaging the parallaxes, as some sources are as close as 100\,pc or as far as 1000\,pc. YSOs with such large deviating distances from expected values near 400\,pc need further investigation, since these can be caused either by uncertainties which are not reflected in $\sigma_\varpi$, or these young stars are not associated with Orion\,A. The combined selection criteria leave us with a final tally of 682 YSOs with IR-excess (23\% of the original YSO catalog) consisting mainly of Class\,II sources (666 Class\,IIs, 16 flat-spectrum) (see Table~\ref{tab:catalog}).
The selected sources have observed $G$ band magnitudes within $\SI{6.3}{mag}<G<\SI{18.2}{mag}$ (see Fig.~\ref{fig:mag_hist}), which is in the range of the suggested magnitude limits \citep{Lindegren2018}\footnote{Bright sources with $G<\SI{6}{mag}$ have generally inferior astrometric quality. Faint sources with $G>\SI{18}{mag}$ are problematic in dense regions.}. As argued above, these sources are the youngest optically visible sources in Orion\,A and hence close to the cloud and a good proxy to the cloud distance.

\section{Results}

In Figure~\ref{fig:mean} we show the average distances, derived from averaged parallaxes of the YSOs per one degree wide bins along Galactic longitude ($\Delta l = \SI{1}{\degree}$, $\SI{\sim7}{pc}$ at $\SI{414}{pc}$, over-sampled by a factor of two). We do not weight the average by the parallax errors, given that we have already applied conservative error cuts. 
The map (Figs.~\ref{fig:scatter} and \ref{fig:mean}, bottom) shows the YSO distribution projected in Galactic coordinates on a Planck-Herschel-Extinction dust column-density map \citep[][hereafter, {\it Herschel map}]{Lombardi2014}\footnote{We use a factor 3,050 to linearly convert $Herschel$ optical depth to extinction, as derived by \citet{Meingast2018}.}.
The distance variations in Fig.~\ref{fig:mean} (top) indicate that the Head of the cloud appears to be roughly on the plane of the sky at about $\SI{400}{pc}$ (for an extent of about $\SI{15}{pc}$ to $\SI{20}{pc}$), while the Tail, starting between $l \approx \SI{210}{\degree}$ and $\SI{211}{\degree}$ and reaching to $l\approx\SI{214.5}{\degree}$, extends from about $\SI{400}{pc}$ to about $\SI{470}{pc}$ along the line-of-sight. Thus, the Tail is inclined $\SI{\sim70}{\degree}$ away from the plane of the sky. Consequently, the  Tail is about four times as long ($\sim$75\,pc) as the Head, leading to a total length of the Orion\,A filament of about 90\,pc. 

The surprising extent of the cloud along the line-of-sight is visualized in Fig.~\ref{fig:xy_stretch}, where we project the YSO positions ($\sigma_\varpi$ as gray scale) in a cartesian plane as seen from the Sun, with $X_{Orion}$ pointing toward Orion\,A. Over-plotted we show the mean positions per bin (orange dots, as in Fig.~\ref{fig:mean}). The displayed mean positions were transformed into the cartesian coordinate system using the following positions: the mean YSO distances ($\bar{d}_\mathrm{YSOs}$), the mean Galactic latitude positions of the YSOs ($\bar{b}_\mathrm{YSOs}$), and the $\Delta l$ bin centers (see also Table~\ref{tab:mean}). 
Sources with $\sigma_\varpi \gtrsim \SI{0.085}{mas}$ disappear in this visualization, while the scatter of YSO distances is still largest near the ONC. However, since the scatter follows largely the line-of-sight, it is still likely that it reflects parallax measurement uncertainties. This should be kept under review in future {\it Gaia} data releases.

In Figure~\ref{fig:xyz} we show the orientation of the Orion\,A cloud projected in Galactic cartesian coordinates, using the mentioned mean YSO positions (Table~\ref{tab:mean}). 
We exclude the three rightmost positions in Fig.~\ref{fig:mean} ($l\leq\SI{208}{\degree}$), since they are not projected on top of high dust column-density. Figure~\ref{fig:xyz} highlights the extent of the cloud in galactic 3D space, also showing an idealized representation of the 3D shape of the cloud in gray scale. The shape is deduced by using extinction contours at $A_\mathrm{K,Herschel} = \SI{0.57}{mag}$ (using extinctions from the {\it Herschel map}). For the far end of the Tail ($l\geq\SI{213.5}{\degree}$, last three points), we extrapolate the cloud shape manually, since the extinction drops on the upper side of the Tail. We use then the middle $b$ position between the upper and lower edge of the Tail, instead of $\bar{b}_\mathrm{YSOs}$. This approach visualizes the opening of the Tail. The sharp turn from Head to Tail is clearly visible in XY and XZ projection. The striking bent of the Head, which consists basically of the Integral Shaped Filament (ISF), calls for a revision of the star-formation history in Orion\,A.  

A potential caveat to using the distances of YSOs as proxies to the cloud distance is that {\it Gaia}, being an optical telescope, is not sensitive to highly extincted sources. As a consequence, it will miss embedded YSOs and non-embedded YSOs that may be hidden by the cloud. 
This implies that the derived distances might suffer from a bias toward closer distances (corresponding to the mean separation between YSOs and the cloud), more pronounced at the denser parts. 
In a first step we tested the average distances by using only sources projected on top of high extinction, by gradually increasing the extinction threshold (from $A_{\mathrm{K,Herschel}} = \SI{0.1}{mag}$ to $\SI{1.0}{mag}$ with \SI{0.1}{mag} steps). Secondly, we used only sources projected on low extinction, again using the mentioned extinction thresholds. We find no significant difference in the mean distance distribution, and also the mean distances do not shift systematically to closer distances at high extinctions. With this, we estimate, given the error in the DR2 parallaxes and the source distance distribution, that the averaged distances per bin are approximately reflecting the cloud shape, especially in regions of low extinction. For regions of higher extinction, like the ISF, the distance might be biased toward closer distances, aggravated by the existence of foreground populations \citep[e.g.,][]{Alves2012,Bouy2014} of young stars. 

We like to point out that in Fig.~\ref{fig:xy_stretch}, the ONC, an especially embedded cluster, appears at about 400 to 410\,pc  \citep[close to 414\,pc,][]{Menten2007} while the adjacent regions (including foreground clusters) appear at a distance of about 390\,pc. The about 10\,pc difference compared to the literature value can be seen as an estimate of remaining systematic uncertainties for the approach we are following. A global systematic parallax offset of 0.08\,mas \citep{Stassun2018} would produce a shift of about 12\,pc toward closer distances at the Head, and of about 16\,pc at the Tail. As mentioned in Sect.~\ref{Data}, we do not include a systematic offset in the reported distances, since it is very unclear how it affects sources across Orion\,A. More importantly for this paper, relative distances are sufficient and the 3D shape of Orion\,A is largely independent of an offset.

We further test the result by a) changing the bin size $\Delta l$ along the cloud from $\SI{0.1}{\degree}$ to $\SI{1.0}{\degree}$, b) varying the different error cuts, and c) excluding sources that are not projected near regions of high dust column-density. The overall result stays the same in all cases, with the Tail starting to incline between $l\approx\SI{210}{\degree}$ and $\SI{211}{\degree}$. Regarding a), using smaller bins naturally increases noise or reflects the existence of cloud sub-structure, while larger bins have a smoothing effect. It is clear from Fig.~\ref{fig:mean}, that, for example, the region near L1641-South shows some significant distance variations, which hint toward a more complex structure than presented here. 
In this paper we will not go into detail about specific sub-structures or sub-clusters in Orion\,A, since we are only interested in the overall shape and line-of-sight extent. A more detailed analysis of this important cloud is called for, using future {\it Gaia} data releases, which will provide improved accuracy. 

In Table~\ref{tab:regions} we provide average distances of large-scale sub-regions in Orion\,A. We find that YSOs at the Head of the cloud, including the ISF region, the ONC, NGC1977, NGC1981, NGC1980, and L1641-North, lie on average at about 395\,pc. YSOs at the Tail are on average at about 430\,pc, including L1641-Center and South, and L1647-North and South. Separating the very southern part (L1647-South), we get a maximum distance to the end of the Tail of about 470\,pc, while the most distant stars have distances of about 550\,pc. 
We find that the two clusters L1647-North and South are more distant (420\,pc to 470\,pc) than estimated with X-ray luminosities \citep[250\,pc to 280\,pc,][]{Pillitteri2016}. To make this a fair comparison, we investigate the DR2 parallaxes of XMM-Newton X-ray sources\footnote{From \url{https://nxsa.esac.esa.int}} in this region, which show a similar average distance as the IR-excess YSOs, supporting the farther distance estimated toward these clusters. The resulting tension between the X-ray luminosities and the {\it Gaia} results need further investigation.

The main result in this paper confirms previous work who pointed out a distance gradient in Orion\,A, as already discussed in Sect.~\ref{Intro}. The $\sim$70\,pc distance difference from Head to Tail is in agreement with \citet{Schlafly2014}, while the individual values along the cloud show variations between the samples (see Tables~\ref{tab:literature}, \ref{tab:mean}, and \ref{tab:regions}). \citet{Kounkel2017} discuss the 3D orientation of Orion\,A using VLBI measurements in the ONC and L1641-South. The $\sim$40\,pc distance difference between these regions is again in agreement with our findings. 
\citet{Kounkel2018}, who also use {\it Gaia} DR2 parallaxes of young stars, find a smaller distance difference from Head to Tail as compared to our result (about 55\,pc from ONC to L1647). The discrepancy is due to the different samples used. 
In this paper we use only the highest-quality data for the youngest YSOs (ages $\lesssim\SI{3}{Myrs}$), as these are likely to be the closest sources to the cloud, while \citet{Kounkel2018} aims at maximizing the identification of young stars in the entire Orion star-forming region, and includes sources as old as 12\,Myrs. 
For completeness, we compare our sample to the \citet{Kounkel2018} sample (K18) and we find that only about 20\% of their sources are in common with our sample (or about 68\% of our sample are in common with K18). The rest of the K18 sources (80\%) are likely older and less connected to the Orion A cloud, hence, not good tracers of the cloud's shape. The sources which are only in our sample (about 1/3 of our sample) are further responsible for the different results. We find that some of these sources are more distant, especially near the Tail.

While these three papers \citep{Schlafly2014,Kounkel2017,Kounkel2018} point to a gradient in the distance from the Head to the Tail of the cloud, our paper not only confirms this gradient, but 1) establishes that the Head of the cloud is bent in regards to the Tail, 2) the Head is essentially on the plane of the sky while the tail appears to be highly inclined, not far from the line-of-sight, and 3) that the cloud has overall a cometary-like shape oriented toward the Galactic plane, although containing sub-structure not resolved in our reconstruction.

Furthermore, our results are in agreement with \citet{Kuhn2018}, who investigate the kinematics of the ONC using {\it Chandra} observed cluster members in combination with {\it Gaia} DR2. They report a distance of about 403\,pc to the ONC (Table~\ref{tab:literature}), similar to the estimated 407\,pc that we find, when looking solely at YSOs near the ONC (Fig.~\ref{fig:xy_stretch}). They point out that the ONC seems to be recessed relative to the immediate surroundings (at $\sim$395\,pc), which we also observe by using IR-excess YSOs \citep[see Figs.~\ref{fig:scatter} or \ref{fig:mean} and figure~21 in][]{Kuhn2018}.

\section{Discussion}

\begin{table*}[!ht] 
\begin{center}
\small
\caption{Mean positions per Galactic longitude bin ($\Delta l$).} 
\begin{tabular}{ccccccccc}
\hline
\hline
  \multicolumn{1}{c}{$\Delta l$ bin center} &
  \multicolumn{1}{c}{$\bar{b}_\mathrm{YSOs}$} &
  \multicolumn{1}{c}{$\bar{d}_\mathrm{YSOs}$} &
  \multicolumn{1}{c}{$X$} &
  \multicolumn{1}{c}{$Y$} &
  \multicolumn{1}{c}{$Z$} &
  \multicolumn{1}{c}{$X_\mathit{Orion}$} &
  \multicolumn{1}{c}{$Y_\mathit{Orion}$} &
  \multicolumn{1}{c}{$Z_\mathit{Orion}$} \\
  
  \multicolumn{1}{c}{(deg)} &
  \multicolumn{1}{c}{(deg)} &
  \multicolumn{1}{c}{(pc)} &
  \multicolumn{1}{c}{(pc)} &
  \multicolumn{1}{c}{(pc)} &
  \multicolumn{1}{c}{(pc)} &
  \multicolumn{1}{c}{(pc)} &
  \multicolumn{1}{c}{(pc)} &
  \multicolumn{1}{c}{(pc)} \\
\hline
  207.0 & -19.20956 & 371.03 $\pm$ 21.83 & -312.18 & -159.06 & -122.08 & 370.22 & -24.44 &  1.60 \\
  207.5 & -19.15597 & 394.01 $\pm$ 30.83 & -330.14 & -171.86 & -129.29 & 393.35 & -22.72 &  2.13 \\
  208.0 & -19.14714 & 396.69 $\pm$ 20.38 & -330.88 & -175.93 & -130.11 & 396.20 & -19.61 &  2.27 \\
  208.5 & -19.24683 & 391.01 $\pm$ 24.00 & -324.42 & -176.15 & -128.89 & 390.68 & -16.10 &  1.61 \\
  209.0 & -19.41924 & 392.69 $\pm$ 25.02 & -323.92 & -179.55 & -130.56 & 392.48 & -12.93 &  0.48 \\
  209.5 & -19.59683 & 393.22 $\pm$ 21.78 & -322.42 & -182.42 & -131.89 & 393.10 &  -9.70 & -0.71 \\
  210.0 & -19.67799 & 390.21 $\pm$ 26.41 & -318.20 & -183.71 & -131.40 & 390.16 &  -6.41 & -1.23 \\
  210.5 & -19.59386 & 395.07 $\pm$ 30.41 & -320.69 & -188.90 & -132.49 & 395.06 &  -3.25 & -0.65 \\
  211.0 & -19.46612 & 401.18 $\pm$ 30.38 & -324.22 & -194.81 & -133.69 & 401.18 &    0.0 &  0.24 \\
  211.5 & -19.36718 & 409.36 $\pm$ 31.97 & -329.28 & -201.79 & -135.75 & 409.34 &   3.37 &  0.94 \\
  212.0 & -19.15993 & 417.43 $\pm$ 43.81 & -334.39 & -208.95 & -137.00 & 417.37 &   6.88 &  2.46 \\
  212.5 & -19.09259 & 423.31 $\pm$ 45.68 & -337.38 & -214.93 & -138.46 & 423.17 &  10.47 &  2.96 \\
  213.0 & -19.22319 & 435.13 $\pm$ 35.87 & -344.59 & -223.78 & -143.27 & 434.89 &  14.34 &  2.02 \\
  213.5 & -19.53701 & 448.16 $\pm$ 32.40 & -352.20 & -233.12 & -149.87 & 447.79 &  18.42 & -0.42 \\
  214.0 & -19.73136 & 461.17 $\pm$ 39.97 & -359.88 & -242.74 & -155.69 & 460.60 &  22.72 & -2.06 \\
  214.5 & -19.74885 & 467.31 $\pm$ 38.40 & -362.47 & -249.12 & -157.90 & 466.54 &  26.85 & -2.30 \\  
\hline
\end{tabular}
\tablefoot{The mean positions per bin ($\Delta l = \SI{1}{\degree}$, within a Galactic latitude range $\SI{-20.5}{\degree}<b<\SI{-18.1}{\degree}$) correspond to the orange dots in Figs.~\ref{fig:mean}, \ref{fig:xy_stretch}, and \ref{fig:xyz}.
The reported mean distances ($\bar{d}_\mathrm{YSOs}$) do not include a systematic global parallax offset. The distance error is the standard deviation of the mean. $XYZ$ are Galactic cartesian coordinates (see also Fig.~\ref{fig:xyz}). $XYZ_\mathrm{Orion}$ are transformed Galactic cartesian coordinates with $X$ pointing toward Orion\,A (see also Fig.~\ref{fig:xy_stretch}).}
\label{tab:mean}
\end{center}
\end{table*}

\begin{table*}[!ht] 
\begin{center}
\small
\caption{Averaged parallaxes and derived distances to different large-scale sub-regions in Orion\,A.} 
\begin{tabular}{lcccccc}
\hline \hline
Region & $l$-Range & Sample  & Mean($\varpi$) & Mean($d$) & Median($\varpi$) & Median($d$) \\
 & (\si{\degree}) & size  &  (mas) & (pc) & (mas) & (pc) \\
\hline
Orion\,A (all)   & 208 -- 215 & 650 & 2.50$\pm$0.20  & 400$\pm$32 & 2.52$\pm$0.10  & 397$\pm$16 \\
Head (ISF)       & 208 -- 211 & 483 & 2.55$\pm$0.16  & 393$\pm$25 & 2.54$\pm$0.08  & 393$\pm$13 \\
Tail             & 211 -- 215 & 145 & 2.33$\pm$0.24  & 428$\pm$42 & 2.33$\pm$0.17  & 430$\pm$31 \\
Tail-L1641       & 211 -- 214 & 130 & 2.36$\pm$0.23  & 424$\pm$42 & 2.35$\pm$0.17  & 426$\pm$31 \\ 
Tail-L1647-South & 214 -- 215 & 15  & 2.14$\pm$0.18  & 467$\pm$32 & 2.17$\pm$0.07  & 461$\pm$15 \\
\hline
\end{tabular}
\tablefoot{The averages per $l$-range are calculated within $\SI{-20.5}{\degree}<b<\SI{-18.1}{\degree}$. The reported parallaxes and distances do not include a systematic global offset. Shown as uncertainties are the standard deviation from the mean and the median absolute deviation. On top of this we expect a systematic error of about 10\,pc.}
\label{tab:regions}
\end{center}
\end{table*}

The 3D shape of Orion\,A, now accessible via the {\it Gaia} measurements, informs and enlightens our knowledge on this fundamental star-formation benchmark. The main result from this work is that Orion\,A is longer than previously assumed and has a cometary shape pointing toward the Galactic plane. Also of note, the Head of the cloud appears to be bent in comparison with the Tail (Fig.~\ref{fig:xyz}). Why would this be the case? One important hint is that the star-formation rate in the Head of the cloud is about an order of magnitude higher than in the Tail (Gro\ss scheld et al., in prep.). Taking this into consideration, one can think of at least two scenarios to explain the enhanced star-formation rate and the shape of the Head: 1) cloud-cloud collision and 2) feedback from young stars and supernovae. Recently, \cite{Fukui2018} interpreted the gas velocities in this region as evidence that two clouds collided about 0.1 Myr ago, and are likely responsible for the formation of the massive stars. While we cannot rule out this scenario with the data presented here, we point out that there is evidence for a young population of foreground massive stars \citep[e.g.,~in NGC\,1980, NGC\,1981,][]{Bally2008,Alves2012,Bouy2014} \citep[cf.][]{Fang2017}, that could provide the feedback necessary to bend the Head of the cloud. An investigation on the second scenario is needed and beyond the scope of this work, but it seems plausible that an external event to the Orion\,A cloud could have taken place in the last million years.

The 3D shape of the cloud clarifies some previous results. For example, \citet{Meingast2018} found evidence for different dust properties in Orion\,A, when comparing the regions in the Head and the Tail of the cloud. They argued, correctly, that the dust in L1641 might not ``see'' the radiation from the massive stars toward the Head of the cloud, and their properties are then not affected by it. Our result validates this view: the dust grains in L1641 lie substantially in the back of the ONC, which contains the most massive stars in the region, and are hence shielded, or too far from the sources of UV radiation.

The deduced length of the Orion\,A filament of 90\,pc makes it similar to the Nessie Classic filament \citep[$\SI{\sim80}{pc}$,][]{Jackson2010}, which is often regarded as a prototypical large-scale filament, or a ``bone" of the Milky Way \citep{Goodman2014}. 
\citet{Zucker2017} undertook an analysis of the physical properties and kinematics of a sample of 45 large-scale filaments in
the literature. They found that these filaments can be distinguished in three broad categories, depending on their aspect ratio and high column-density fraction. Orion\,A has an average aspect ratio of about 30:1 when taking the length of 90\,pc and its average width (FWHM $\sim$3\,pc), and a high-column-density fraction of about 45\%. For the latter we use an A$_K$ threshold of 0.5\,mag, comparable to  $1\times10^{22}\mathrm{cm}^{-2}$ in \citet{Zucker2017}. This puts Orion\,A squarely into their category c), which describes highly elongated, high-column-density filaments, or so called "bones" of the Milk Way. The position-angle between Orion\,A and the plane is in agreement with the average position-angles of the bones in their sample, but Orion\,A differs significantly from the known bones regarding its distance from the mid-plane of the Milky Way ($\sim$145\,pc), which is an order of magnitude larger than the median distance between bones and the Galactic plane. This discrepancy calls for a large-scale process to have pushed the cloud this far from the plane. \cite{Franco1986} proposed a scenario for the origin of the Orion complex as the consequence of an impact of a high-velocity cloud with the plane of the Galaxy (from above) that could account for the abnormal distance of Orion below the plane. Nevertheless, the cloud's cometary shape with a star-bursting Head closer to the plane, as shown in this work, seems to be at odds with this scenario.

Finally, we point out that the unexpected length of Orion\,A along the line-of-sight affects the observables toward the cloud (masses, luminosities, binary separations) that will need revision. For example, the current cloud and YSO masses toward the Tail can be underestimated by about 30\% to 40\% under the common assumption of a single constant distance to Orion A, while binary separations can be underestimated by about 10\% to 20\%.

\section{Summary}
We have used the recent {\it Gaia} DR2 to investigate the 3D shape of the Orion\,A GMC.
Orion\,A is not the straight filamentary cloud that we see in (2D) projection, but instead a cometary-like cloud, oriented toward the Galactic plane, with two distinct components: a denser and enhanced star-forming (bent) Head, and a lower density and star-formation quieter $\sim$75 pc long Tail. The two components seem to overlap between $l\approx\SI{210}{\degree}$ to $\SI{211}{\degree}$.
We find that the Head of the Orion\,A cloud appears to be roughly on the plane of the sky (at $\sim$400\,pc), while the Tail, surprisingly, appears to be highly inclined, not far from the line-of-sight ($\SI{\sim 70}{\degree}$), reaching at least $\sim$470\,pc. 
The true extent of Orion\,A is then not the projected $\sim$40\,pc but $\sim$90\,pc, making it by far the largest molecular cloud in the local neighborhood. Its aspect ratio ($\sim$30:1) and high-column-density fraction ($\sim$45\%) make it similar to large-scale Milky Way filaments (bones), despite its distance to the galactic mid-plane being an order of magnitude larger than typically found for these structures.
\textit{Gaia} is opening an important new window in the study of the ISM, in particular the star-forming ISM, bringing the critical third spatial dimension that will allow not only cloud structure studies similar to the ones presented here, but an unique view on the dynamics between dense gas and YSOs.

\begin{acknowledgements}
We thank the anonymous referee whose comments helped to improve the manuscript.
J.\,Gro\ss schedl acknowledges funding by the Austrian Science Fund (FWF) under project number P 26718-N27.
This work is based on observations made with ESO Telescopes at the La Silla Paranal Observatory under program ID 090.C-0797(A).
This work is part of the research program VENI with project number 639.041.644, which is (partly) financed by the Netherlands Organisation for Scientific Research (NWO). A.\,Hacar thanks the Spanish MINECO for support under grant AYA2016-79006-P.
J.\,Alves is part of the Research Platform Data Science @ Uni Vienna (\url{https://datascience.univie.ac.at}).
This work has made use of data from the European Space Agency (ESA) mission {\it Gaia} (\url{https://www.cosmos.esa.int/gaia}), processed by the {\it Gaia} Data Processing and Analysis Consortium (DPAC, \url{https://www.cosmos.esa.int/web/gaia/dpac/consortium}). Funding for the DPAC has been provided by national institutions, in particular the institutions participating in the {\it Gaia} Multilateral Agreement. This research has made use of the VizieR catalog access tool, CDS, Strasbourg, France. This research has made use of Python, \url{https://www.python.org}, of Astropy, a community-developed core Python package for Astronomy \citep{Astropy2013}, NumPy \citep{Walt2011}, and Matplotlib \citep{Hunter2007}. This research made use of TOPCAT, an interactive graphical viewer and editor for tabular data \citep{Taylor2005}.
This work has made use of ``Aladin sky atlas'' developed at CDS, Strasbourg Observatory, France \citep{Bonnarel2000, Boch2014}.
\end{acknowledgements}

% BIBLIOGRAPHY
\begin{flushleft}
\bibliographystyle{aa}
\bibliography{biblio} 
\end{flushleft}

% APPENDIX
\clearpage  
\onecolumn
\begin{appendix}
\section{Additional figure}
\begin{figure*}[!ht]
    \centering
    \includegraphics[width=1\linewidth]{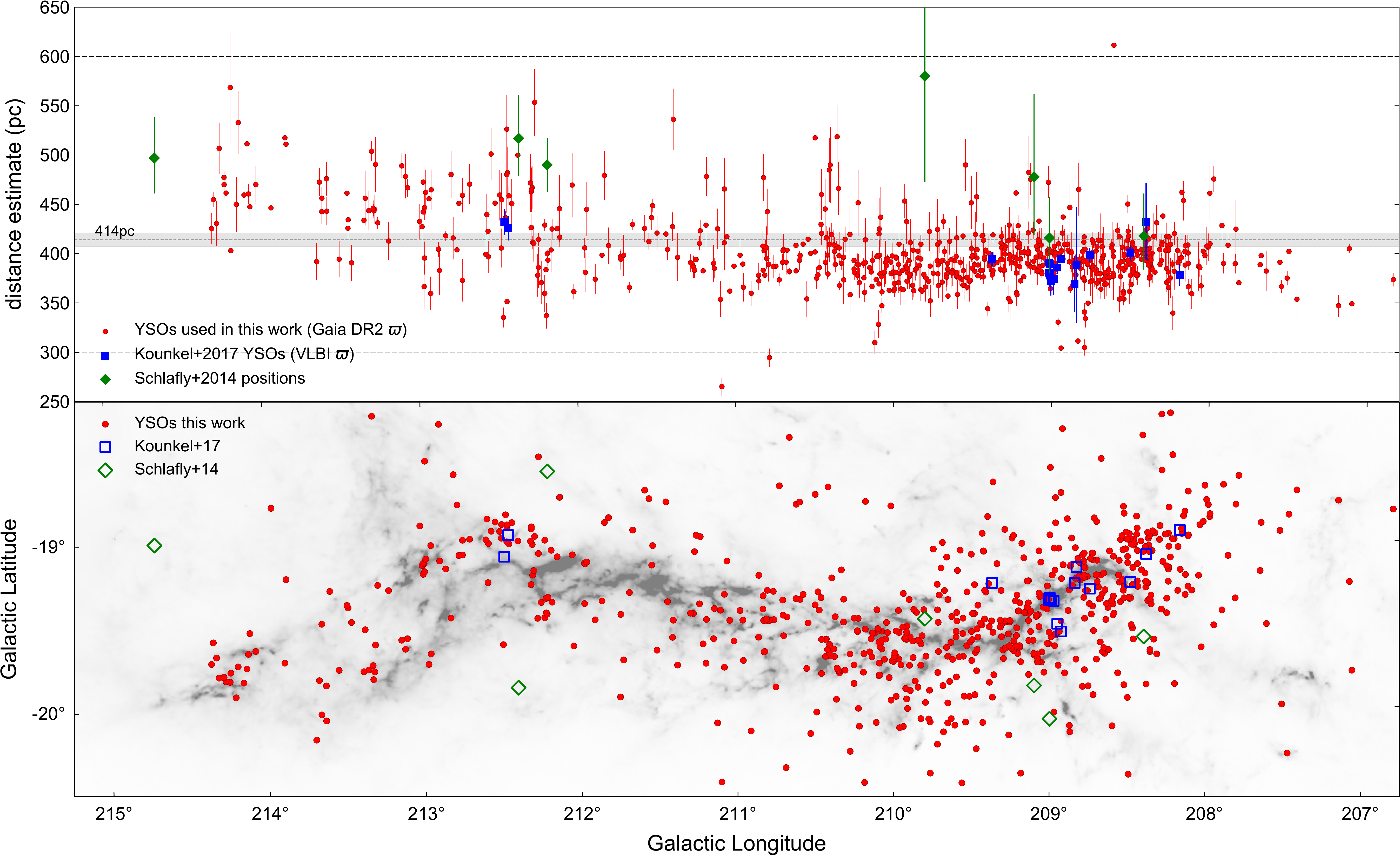}
    \caption{YSO distribution and distance estimates toward Orion\,A, similar to Figs.~\ref{fig:scatter} and \ref{fig:mean}. Additionally we show the positions of measured distances from the Literature. The gray band shows the distance of 414$\pm$7\,pc from \citet{Menten2007}. Blue boxes are the reported distances from \citet{Kounkel2017} (VLBA parallaxes). Green diamonds show distances to certain positions as given in \citet{Schlafly2014} (optical reddening). The reported distances from previous works are largely in agreement with {\it Gaia} DR2 distances of YSOs, within the errors and the scatter. See also Table~\ref{tab:literature}.}
    \label{fig:scatter_literature}
\end{figure*}

\section{YSO table} \label{catalog}

\begin{table*}[!ht] 
\begin{center}
\small
\caption{Catalog of the 682 YSOs, used to infer on the cloud's shape.} 
\label{tab:catalog}
\begin{tabular}{cccccc}
\hline \hline
  \multicolumn{1}{c}{{\it Gaia} DR2 source\_id} &
  \multicolumn{1}{c}{RAJ2015.5} &
  \multicolumn{1}{c}{DEJ2015.5} &
  \multicolumn{1}{c}{$\varpi$\tablefootmark{a}} &
  \multicolumn{1}{c}{$\sigma_\varpi$\tablefootmark{a}} &
  \multicolumn{1}{c}{Class\tablefootmark{b}} \\
 & (h:m:s) & (d:m:s) & (mas) & (mas) & \\
\hline
  3011883130996177280 & 05:42:00.09 & -10:01:11.35 & 2.222 & 0.137 & II\\
  3011892137543646080 & 05:43:27.01 & -09:59:37.67 & 2.199 & 0.038 & II\\
  3011892790378687744 & 05:42:59.94 & -10:03:40.57 & 2.351 & 0.088 & II\\
  3011893408853983232 & 05:42:37.10 & -10:03:29.98 & 1.973 & 0.102 & II\\
  3011893786811104000 & 05:42:34.89 & -10:01:46.50 & 2.127 & 0.061 & II\\
\hline
\end{tabular}
\tablefoot{
Only the first five rows are given. The full table is available in electronic form at the CDS.
\tablefoottext{a}{The parallax ($\varpi$) and its error ($\sigma_\varpi$) are given. Further {\it Gaia} parameters can be obtained at the {\it Gaia} Archive (\url{https://gea.esac.esa.int/archive/}), using the {\it Gaia} DR2 source\_id for cross-matching.}
\tablefoottext{b}{YSO classification: Class\,II (II), flat-spectrum source (F).}
}
\end{center}
\end{table*}

\end{appendix}
\end{document}